\documentstyle[prl,aps,twocolumn,epsf]{revtex}
\begin{document}
{\bf Comment on "Peierls Gap in Mesoscopic Ring Threated by a Magnetic
Flux"} \bigskip

Yi {\it et al.}\cite{Yi97} have recently considered
the stability of a Charge Density Wave (CDW) in a clean mesoscopic 1D ring
pierced by an Aharonov-Bohm
flux. Although this letter rises very interesting questions, some results
are incorrect or incomplete.

The main result is that a threading flux tends to suppress the Peierls
instability, as also claimed in another recent work\cite{Visscher96}.
This interesting
result is only partly correct because it does not properly take into account
the parity effect essential in a 1D ring: the thermodynamics depends
crucially on the parity of the number $N_e$ of electrons (forgetting the
spin)\cite{Cheung89,Montambaux97}.

The stability of the  CDW is studied through the calculation of the
polarization function $\chi$ which, in a finite system, is a discrete
sum where the wave vector and
the nesting vector can take only quantized values. To account for the finite
size, the nesting vector is indeed quantized in ref.\cite{Yi97} but the sum
is calculated as an integral which  does not exhibit the parity
effect.
The sum (4) of ref.\cite{Yi97} can be indeed calculated exactly. At the
best nesting vector $q = 2 k_F$, one
gets, using the same notations as in ref.\cite{Yi97}:  $$\chi_{2 k_F} ={m R
\over  N_e \hbar^2
k_F} \left(\psi(2 k_F R)- {1 \over 2}[ \psi (| f |) + \psi (1-| f
|)] \right)$$
for an even number $N_e$ of electrons and $-1/2<f<1/2$.
$f$ is the dimensionless flux
$\phi/\phi_0$. $\psi$ is the digamma function. $\chi_{2k_F}$ does not vary
logarithmically when $f \rightarrow 0$, as claimed in
ref.\cite{Yi97},
but as a power law. The critical flux $f_c$
 does not vary
linearly with the size as claimed in ref.\cite{Yi97}.

More important,
 the limits of the discrete sum depend on the parity,
 as mentioned in the footnote [11] of ref.\cite{Yi97}.
Performing the same summation when $N_e$ is odd gives a similar expression
for $\chi_{2 k_F}$ as above where
$f$ is
changed into
$f-1/2$. Consequently, for a small ring, the Peierls phase does not exist
for zero flux and
it is {\it stabilized} above the critical flux $f'_c=1/2 -f_c$.

More generally, the variation of the
order parameter $\Delta$ is given by a general equation of the form
$\chi(\Delta,T)=1/Cte$ where the constant is proportional to the interaction
 parameter.
The generalized polarization function  $\chi(\Delta,T)$ has the structure
of a discrete sum
$\sum_n F(n+f)$ which is periodic. The limits of the sum depend on the parity.
	Using the Poisson summation formula, this sum can be replaced by an
integral plus an harmonic expansion in $f$:
$\chi=\int_{-\infty}^{\infty}  F(y) dy + 2 \sum_{m>0} G_m \cos 2 \pi m f$
where  $G_m$ has the sign of
$(-1)^{N_e m}$. One immediately deduces that changing the parity is equivalent
to a shift of the flux by half a period.
The complete dependence of the critical temperature and of the gap with the flux and the size
are
calculated in ref.\cite{Montambaux97}.

Two other comments are of importance concerning the persistent current.
First,
it is true that the current is {\it a priori} weak in the Peierls phase and
recovers its
metallic value above the critical flux $f_c$ (for even $N_e$). However, the CDW gap vanishes
continuously at the critical flux (as agreed by the authors on their
fig.1), so that the current cannot
be discontinuous at $f_c$. Indeed the current, very weak at small flux,
increases with the flux due to the decrease of the gap and varies {\it
continuously} at $f_c$. It is found to vary almost linearly below
 $f_c$\cite{Montambaux97} (fig.1a).
Secondly,
 The current
shown on the fig.2b of ref.\cite{Yi97} does not present the correct parity
effect: the slope has to be always negative in the normal phase, whatever
the  parity\cite{Cheung89}.
Indeed, when $N_e$ is odd, the CDW does not exist for small rings and it 
is {\it restored} above the critical flux
$f'_c=1/2-f_c$. Thus, the current for both parities are simply shifted by
half a period (fig.1b).
\begin{figure}[hbt]
\centerline{\epsfxsize 6cm \epsffile{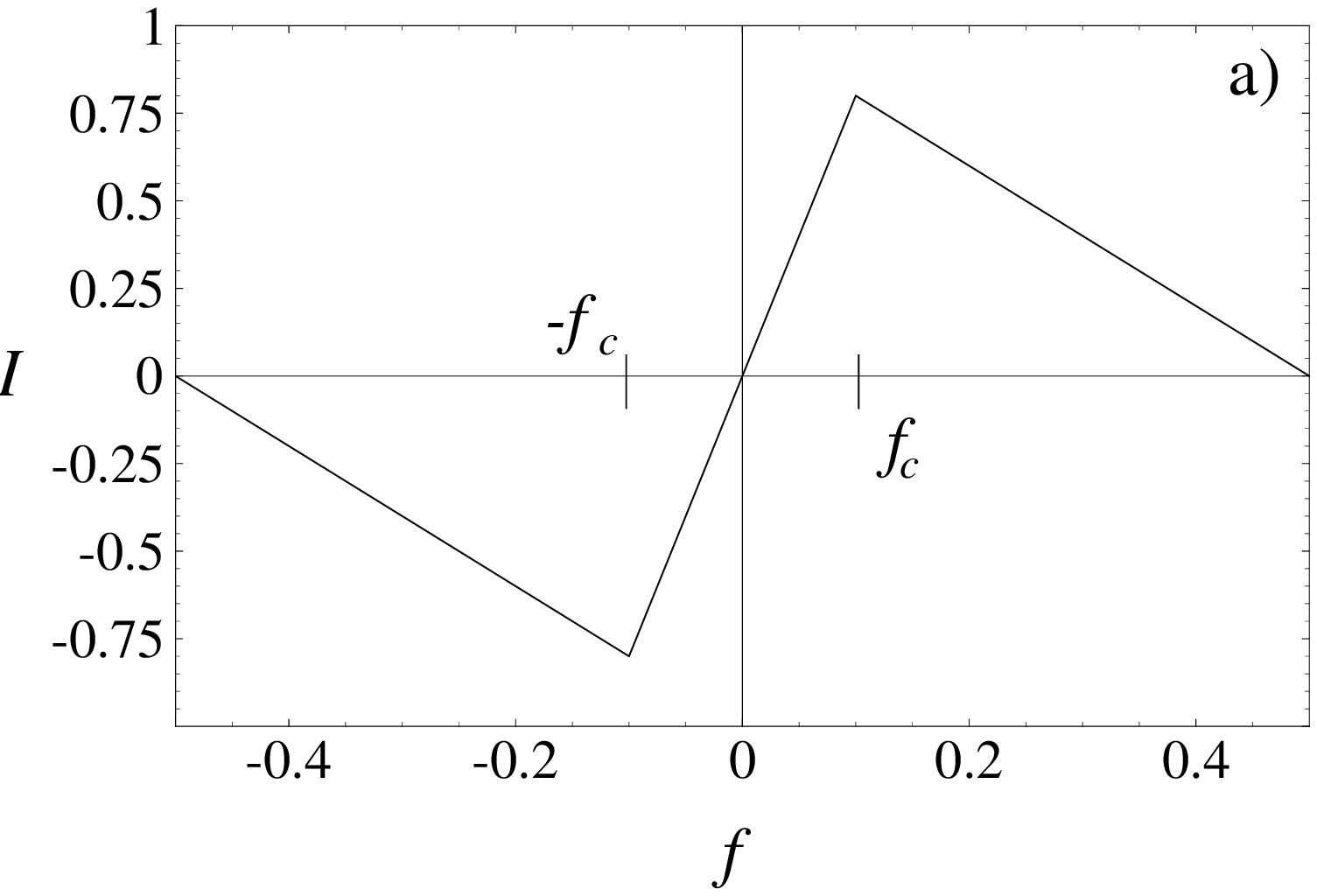}}
\centerline{\epsfxsize 6cm \epsffile{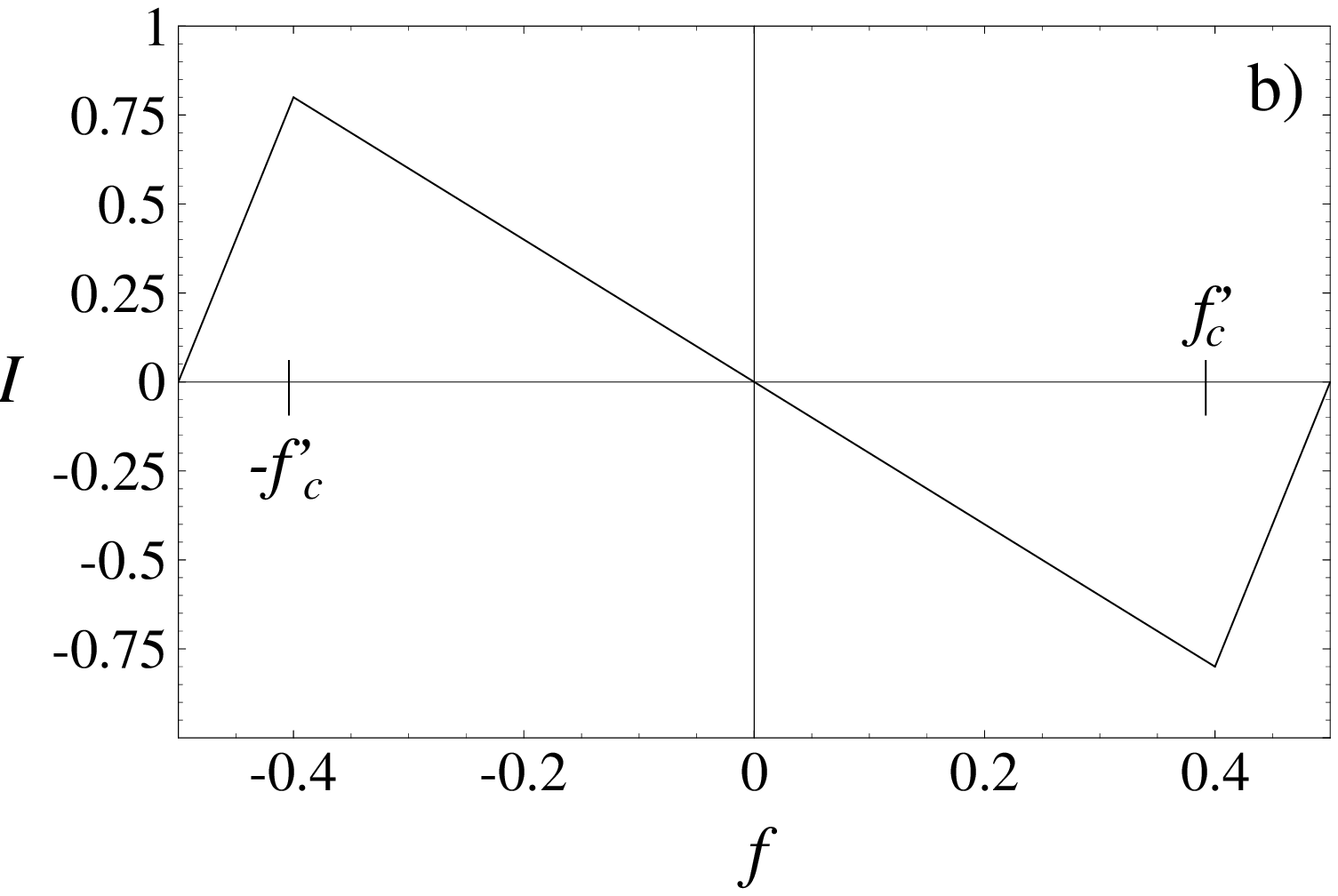}}
\caption{Persistent current (a) for  even $N_e$,
(b) for odd $N_e$.
} \end{figure}

{\small G. Montambaux,

Laboratoire de Physique des Solides,  associ\'e au
CNRS

 Universit\'{e} Paris--Sud
 91405 Orsay, France}

\end{document}